\documentclass[10pt,conference]{IEEEtran}

\IEEEoverridecommandlockouts
\usepackage{cite}
\usepackage{amsmath,amssymb,amsfonts}
\usepackage{enumitem}
\usepackage{algorithmic}
\usepackage{graphicx}
\usepackage{textcomp}
\usepackage{xcolor}
\def\BibTeX{{\rm B\kern-.05em{\sc i\kern-.025em b}\kern-.08em
    T\kern-.1667em\lower.7ex\hbox{E}\kern-.125emX}}

\usepackage{caption}
\usepackage{float}
\usepackage{subfigure}
\usepackage{tikz}
\usetikzlibrary{arrows,%
	shapes,positioning}
\usepackage{color,soul}
\usepackage{booktabs}
\usepackage{algorithm}
\usepackage{mathtools}
\usepackage[T1]{fontenc}
\usepackage{isomath} 
\usepackage{txfonts}
\usepackage{upgreek} 
\usepackage{array}
\usepackage{multirow}
\usepackage{verbatim}
\usepackage{nicefrac}

\usepackage{hyperref}
\usepackage{setspace}

\newcolumntype{P}[1]{>{\centering\arraybackslash}p{#1}}
\newcolumntype{M}[1]{>{\centering\arraybackslash}m{#1}}

\newcommand*{\QEDB}{\hfill\ensuremath{\square}}

\newcommand{\Rest}{{\hat R}}
\newcommand{\Gapprox}{{\tilde G}}

\newcommand\blfootnote[1]{%
  \begingroup
  \renewcommand\thefootnote{}\footnote{#1}%
  \addtocounter{footnote}{-1}%
  \endgroup
}

\begin{document}

\title{Perturb and Combine to Identify Influential Spreaders in Real-World Networks}

\author{\IEEEauthorblockN{Antoine J.-P. Tixier}
\IEEEauthorblockA{\'Ecole Polytechnique \\
Palaiseau, France}
\and
\IEEEauthorblockN{Maria Evgenia G. Rossi}
\IEEEauthorblockA{\'Ecole Polytechnique \\
Palaiseau, France}
\and
\IEEEauthorblockN{Fragkiskos D. Malliaros}
\IEEEauthorblockA{Centrale-Sup\'elec \\
Gif-sur-Yvette, France}
\and
\IEEEauthorblockN{Jesse Read}
\IEEEauthorblockA{\'Ecole Polytechnique \\
Palaiseau, France}
\and
\IEEEauthorblockN{Michalis Vazirgiannis}
\IEEEauthorblockA{\'Ecole Polytechnique \\
Palaiseau, France}
}

\maketitle


\begin{abstract}
\blfootnote{Accepted as long paper (oral presentation) at ASONAM 2019.}
Some of the most effective influential spreader detection algorithms are unstable to small perturbations of the network structure.
Inspired by bagging in Machine Learning, we propose the first Perturb and Combine (P\&C) procedure for networks. It (1) creates many perturbed versions of a given graph, (2) applies a node scoring function separately to each graph, and (3) combines the results. Experiments conducted on real-world networks of various sizes with the $k$-core, generalized $k$-core, and PageRank algorithms reveal that P\&C brings substantial improvements. Moreover, this performance boost can be obtained at almost no extra cost through parallelization. Finally, a bias-variance analysis suggests that P\&C works mainly by reducing bias, and that therefore, it should be capable of improving the performance of all vertex scoring functions, including stable ones. 
\end{abstract}


\section{Introduction}

Influential spreaders are nodes that can diffuse information to the largest part of the network in a minimum amount of time. Detecting influential spreaders is an important task with numerous real-world applications. Well known examples include epidemiology \cite{hoppensteadt1975mathematical}, viral marketing \cite{leskovec2007dynamics}, social media analysis \cite{pei2014searching}, expert finding \cite{balog2006formal}, and keyword extraction \cite{tixier2016graph}.

\begin{figure}[h]
    \centering
    \captionsetup{size=small}
	\includegraphics[width=.65\columnwidth]{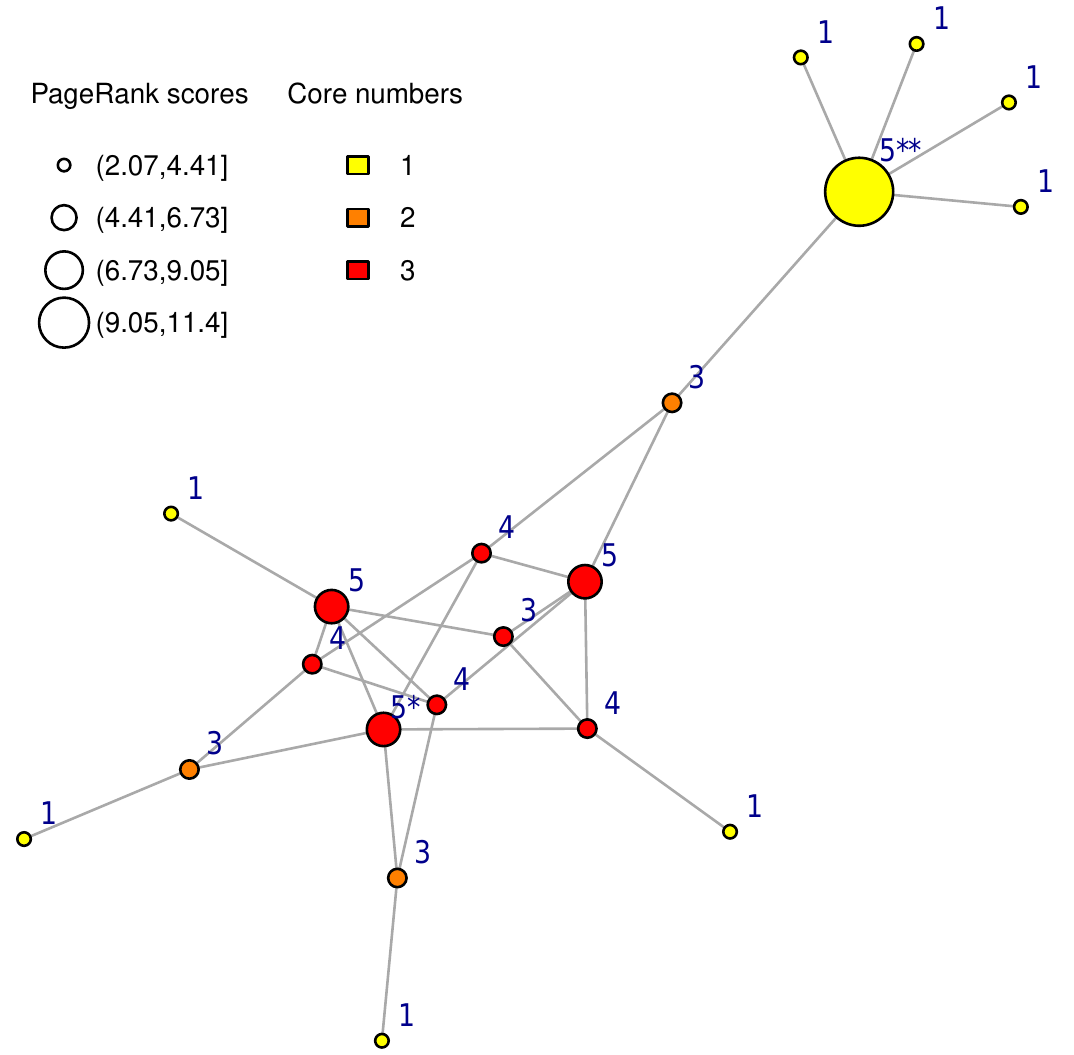}
	\setlength{\belowcaptionskip}{-18pt}
	\caption{Degree vs PageRank vs $k$-core. Labels are degrees. \label{fig:deg_vs_core}}
\end{figure}

Intuitively, one could think that nodes with many connections are the most influential spreaders, and use node centrality criteria such as the degree or PageRank as a measure of influence. However, there are cases when node degree is not a good spreading predictor. Consider for example a hub node located at the periphery of the network. As shown in Fig. \ref{fig:deg_vs_core}, nodes $*$ and $**$ both have same degree (5) and high PageRank scores (resp. in $(6.73,9.05]$ and $(9.05,11.4]$). However, node $*$ lies in a much more central location and is therefore a much better spreader, which is captured by its higher core number (3 vs 1) but not by degree or PageRank (the PageRank score of node $**$ is even greater than that of node $*$).

As they inherently capture cohesiveness, \textit{graph degeneracy} algorithms are very effective at locating influential spreaders (and better than degree and PageRank) \cite{kitsak2010identification,pei2014searching,de2014role,malliaros2016locating}. Their efficiency, combined with their low time complexity, has motivated a new line of research  \cite{malliaros2019core}. 

\begin{figure}[ht]
    \centering
    \captionsetup{size=small,belowskip=0pt}
	\includegraphics[width=.5\columnwidth]{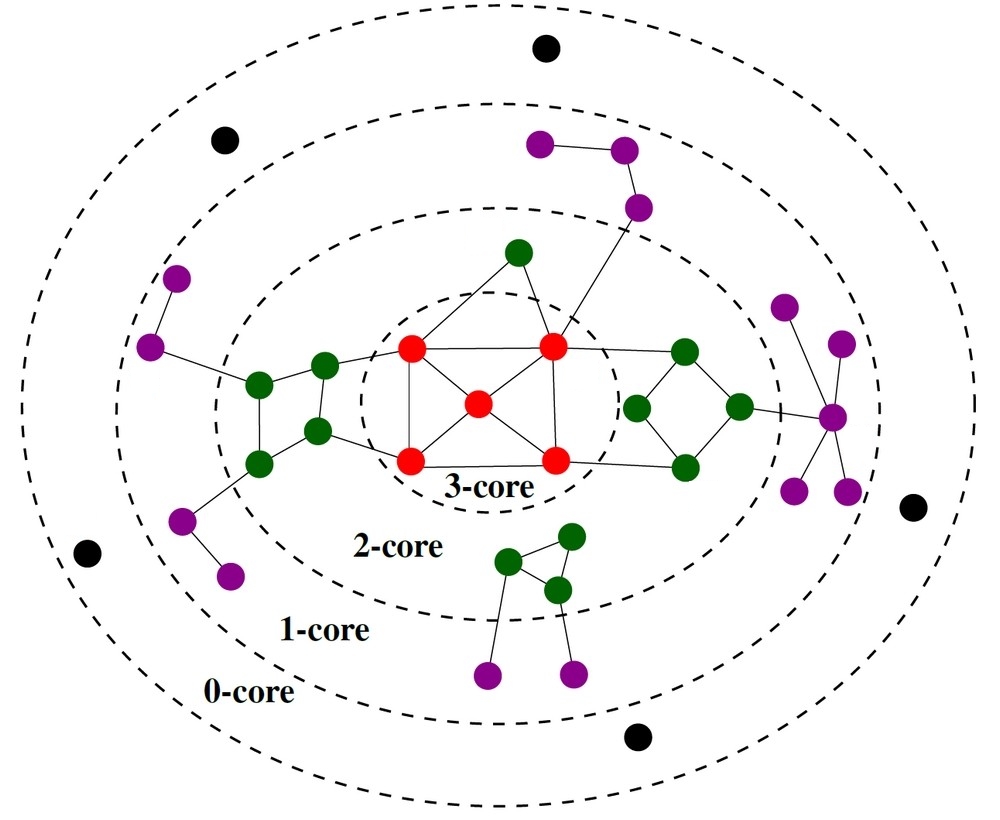}
	\setlength{\abovecaptionskip}{-2pt}
	\setlength{\belowcaptionskip}{-7pt}
	\caption{Illustration of the $k$-core decomposition\label{fig:kcore}. Here, $k_{max}=3$.}
\end{figure}

The most famous member of the graph degeneracy family is the $k$-core algorithm. A $k$-core of a graph $G$ is defined as the maximal subgraph of $G$ in which every vertex $v$ has at least degree $k$ \cite{seidman1983network}. As shown in Fig. \ref{fig:kcore}, the $k$-core decomposition of $G$ is the set of all its cores from 0 ($G$ itself) to $k_{max}$ (its main core). It forms a hierarchy of nested subgraphs whose cohesiveness increase with $k$. A node has core number $k$ if it belongs to a $k$-core but not to a $(k+1)$-core.

\vspace{-0.45cm}

\noindent The basic $k$-core algorithm has time complexity linear in the number of edges, but it is \textit{unweighted}, i.e., it uses as the degree of $v$ the count of its neighbors. A \textit{generalized} version has been proposed by \cite{batagelj2002generalized} for any local monotone vertex property function, with still very affordable time complexity $\mathcal{O}(|E|\log(|V|))$. By using the sum of the weights of the incident edges as the vertex property function, we obtain the \textit{weighted} $k$-core decomposition.

However, graph degeneracy algorithms are unstable to small perturbations of the network structure, such as the removal of a small fraction of edges from the network \cite{adiga2013robust,goltsev2006k}. This opens the gate to improvement: indeed, it is well known in Machine Learning that the performance of unstable algorithms can be improved by using Perturb and Combine (P\&C) strategies.

Therefore, inspired by the bootstrap aggregating (bagging) method, we propose a procedure that first creates many perturbed versions of a given graph, then applies a node scoring function separately to each graph, and finally aggregates the results back into more robust scores. Our contributions are fourfold:

(1) we propose what is, to the best of our knowledge, the first application of P\&C to networks,
(2) experiments on large social networks and small word co-occurrence networks reveal that for $k$-core, generalized $k$-core, and PageRank, P\&C allows to identify better spreaders,
(3) our procedure is trivially parallelizable, so performance gains can be obtained at little extra cost,
(4) we explain through a theoretical analysis why our P\&C strategy is effective. We notice that improvement comes mainly from reducing bias, which implies that P\&C for networks could work well for all algorithms, including stable ones.

\section{Perturb and Combine in Machine Learning}\label{sub:bagging}

In Machine Learning, unstable algorithms are algorithms for which small changes in the training set result in large changes in predictions. These models are also known as \textit{low bias-high variance algorithms} or \textit{strong learners} \cite{breiman1996bias}. Decision trees (especially deep, unpruned ones) in classification and regression are good examples of such models. Indeed, adding or removing only a few observations to/from the training set changes the structure, and thus the predictions, of decision trees.

It is well known that unstable learners can have their accuracy greatly improved by perturbing and combining. Of all the P\&Cs strategies, \textbf{b}ootstrap \textbf{agg}regat\textbf{ing} (bagging) \cite{breiman1996bagging} is certainly the most popular, and one of the most effective. It is actually one of the key ingredients of the acclaimed Random Forest model \cite{breiman2001random}.\\
In the context of decision trees, bagging simply consists in training unpruned trees in parallel on bootstrap samples of the training set. Each bootstrap sample is a \textit{perturbed version} of the training set which is generated by drawing from it with replacement until a set of the same size is obtained. To issue a forecast for a new observation, the predictions of all the trees are combined, e.g., through averaging in regression and majority voting in classification. Note that instability is mandatory for bagging to function well. For instance, bagging is not effective with nearest-neighbors approaches \cite{breiman1996bagging}, which are very stable.

\section{Perturb and Combine for Networks}\label{sec:our_approach}

\subsection{Key idea}
\begin{figure}[H]
    \centering
	\subfigure{\fbox{\includegraphics[width=0.75\textwidth]{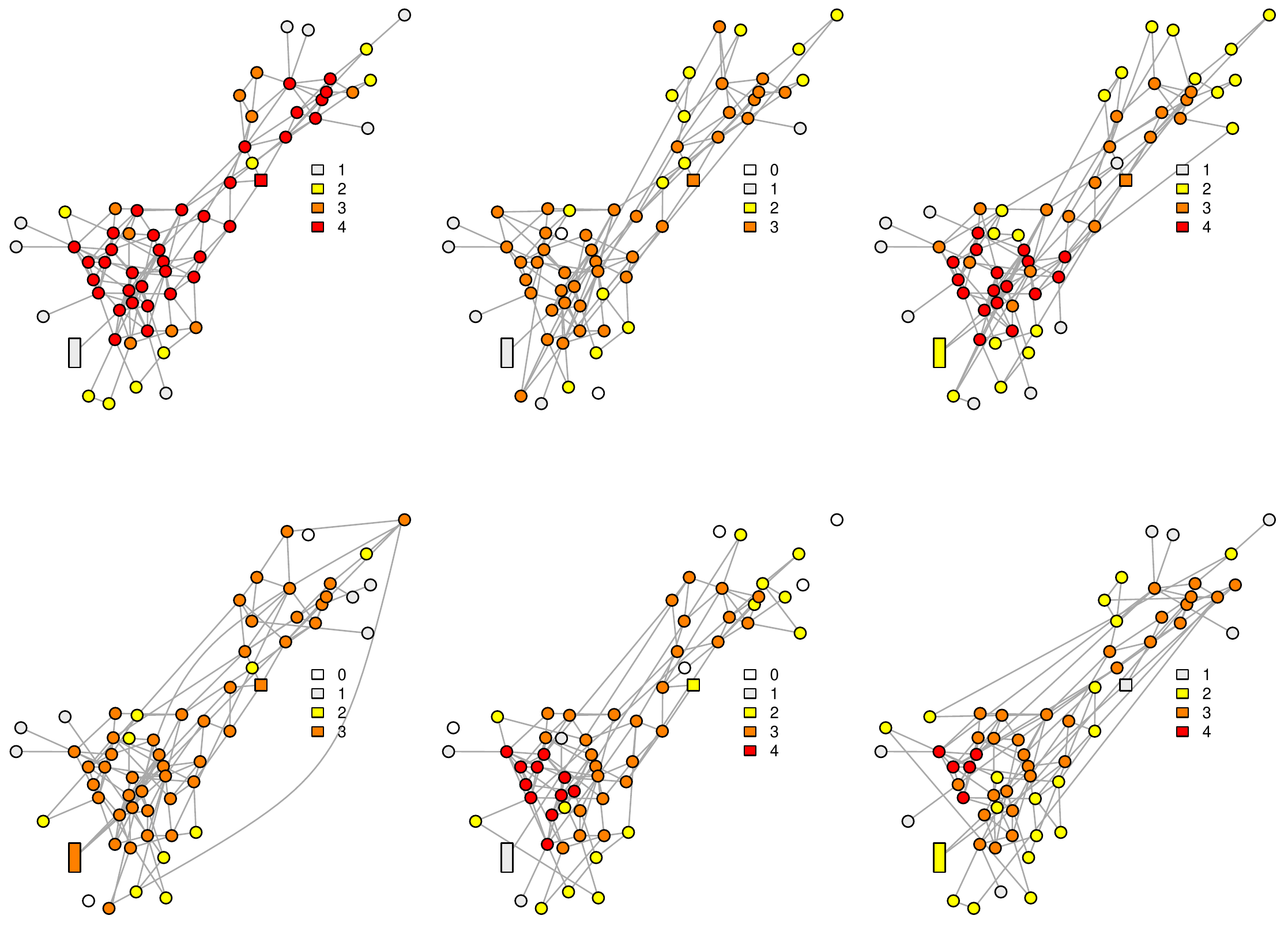}}}

	\subfigure{\fbox{\includegraphics[width=0.75\textwidth]{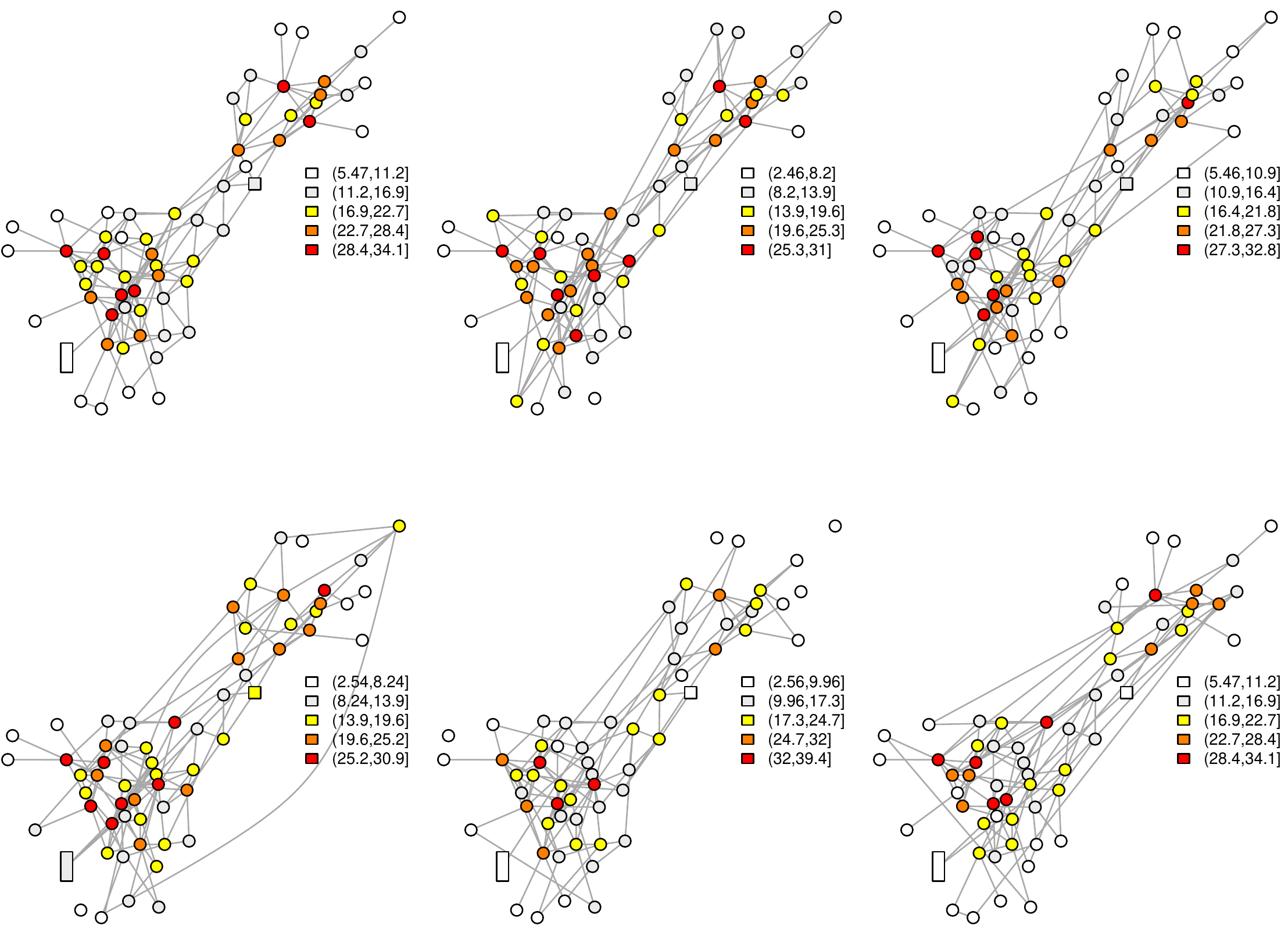}}}
	\captionsetup{size=small}
	\caption{\textbf{Top:} original graph (upper left corner) and 5 perturbed versions of it (generated by the edge perturbation strategy presented in subsection \ref{sub:perturb} with $\varepsilon_a=0.1,\varepsilon_d=0.3,\mathbb{G}=ER$, and $\delta_{w}=0$). Node colors indicate unweighted $k$-core numbers. \textbf{Bottom:} same, but with PageRank. For $k$-core, we observe quite some variability in the core numbers across the different versions of the graph: the average cosine similarity between the 6 rankings is 93.29\%. This illustrates well the unstable nature of the $k$-core vertex scoring function. PageRank rankings are only slightly more stable (average cosine similarity of 94.54\%). The real-world network used in this example is the well-known \textit{dolphins} network \cite{lusseau2003emergent}.\label{fig:per}}
\end{figure}

It was shown that the core decomposition of a network is highly sensitive to small perturbations at the edge level such as edge addition and deletion \cite{adiga2013robust}. This property is easy to verify in practice, as illustrated by Fig. \ref{fig:per} in App. C. \cite{goltsev2006k} also demonstrated that the removal of only a tiny fraction of vertices can vastly damage the $k$-core.

Motivated by these observations, we posit that like unstable learners in Machine Learning, \textbf{\textit{degeneracy-based node scoring functions, and more generally any unstable node scoring function, benefits from using a P\&C strategy}}.

That is, it is possible to identify better spreaders by aggregating node scores computed on multiple perturbed versions of the original network rather than by using the scores computed on the original network. We therefore propose a P\&C procedure for networks described by Algorithm \ref{alg:bagging}, that features three simple steps:

\begin{enumerate}
    \item create $M$ perturbed versions of a given graph,
    \item separately apply a node scoring function to each graph,
    \item aggregate the results.
\end{enumerate}

\begin{algorithm}[]
{\small
\caption{\textsc{PC-NET (P\&C for NETWORKS)}\label{alg:bagging}}
 \begin{algorithmic}[1]
  \REQUIRE original graph $G(V,E)$, $M \in \mathbb{N}$, vertex scoring function $s: V \mapsto \mathbb{R}$ 
  \ENSURE P\&C scores $s_{\mathrm{pc}}(1), \dots ,s_{\mathrm{pc}}(n)$ for the $n$ nodes in $V$ ($n=|V|$)
  \STATE $B \leftarrow$ empty $M \times n$ array
  \FOR{$m \in \big[1, \dots , M\big]$}
\STATE $\tilde{G}_m \leftarrow$ \textsc{PERTURB}($G$)
\STATE $B\big[m,:\big] \leftarrow$ \textsc{MINE}($\tilde{G}_m,s$)
\ENDFOR
\STATE $s_{\mathrm{pc}}(1), \dots ,s_{\mathrm{pc}}(n) \leftarrow$ \textsc{COMBINE}($B$)
\STATE \textbf{return} $s_{\mathrm{pc}}(1), \dots ,s_{\mathrm{pc}}(n)$
\end{algorithmic}
}
\end{algorithm}

\noindent Perturbation models have been widely used for generating graphs, describing them, and studying their behavior. However, this study is, to the best of our knowledge, the first attempt at using graph perturbation not only for descriptive or generative purposes, but as the first step of a process subsequently involving \textit{scoring} and \textit{combination}.

Looking at Algorithm \ref{alg:bagging}, it appears immediately that like Random Forest, \textsc{PC-NET} is a meta-algorithm that can trivially be parallelized (\texttt{for} loop in lines 2-5). The P\&C scores therefore do not take more time to obtain than the original scores, provided that $M$ workers are available. The only additional cost comes from the \textsc{PERTURB} step in line 3, but it can be implemented efficiently. Details about the \textsc{PERTURB}, \textsc{MINE}, and \textsc{COMBINE} parts (lines 3, 4, and 6) are provided next, in subsections \ref{sub:perturb}, \ref{sub:mine}, and \ref{sub:combine}.

\subsection{Perturb}\label{sub:perturb}

\noindent \textbf{High-level framework}. We used a flexible framework very similar to that used in \cite{adiga2013robust}. It is a general model for edge-based perturbation of which most of the perturbation models found in the literature can be seen as special cases. Not using node-based perturbation makes combination easier, since each node is assigned a score in each perturbed graph.

\noindent Let $G(V,E)$ be the original graph and $\mathbb{G}$ be a random graph model. The corresponding perturbation model $\Theta(G, \mathbb{G}, \varepsilon_a, \varepsilon_d)$ is defined as:

\begin{equation}
	\mathbb{P}_{\Theta}\big[(u,v)\big] = \left\{
	\begin{array}{lr}
		\varepsilon_a\mathbb{P}_{\mathbb{G}}\big[(u,v)\big] & \text{if}~(u,v) \notin $E$\\
		\varepsilon_d\mathbb{P}_{\mathbb{G}}\big[(u,v)\big] & \text{if} ~(u,v) \in $E$ \\
	\end{array}
	\right.
\end{equation}

\noindent where $\mathbb{P}_{\Theta}\big[(u,v)\big]$ is the probability of
adding/deleting the edge $(u,v)$, $\mathbb{P}_{\mathbb{G}}\big[(u,v)\big]$ is the probability of selecting the edge $(u,v)$ according to the random graph model $\mathbb{G}$, and $\varepsilon_a$, $\varepsilon_d$ are the probabilities of edge addition and deletion, respectively. 

An interpretation can be given as follows: if the edge $(u,v)$ already exists, then it is deleted with probability
$\varepsilon_d\mathbb{P}_{\mathbb{G}}\big[(u,v)\big]$, and if not, it is added with probability $\varepsilon_a\mathbb{P}_{\mathbb{G}}\big[(u,v)\big]$. By XOR-ing the original graph $G$ with one realization $\theta \sim \Theta(G, \mathbb{G}, \varepsilon_a, \varepsilon_d)$ of the perturbation model, we obtain the perturbed graph $\tilde{G}=G \oplus \theta$. Depending on the random graph model $\mathbb{G}$ used, we obtain a different perturbation scheme.

\noindent \textbf{Edge weight awareness}.
To make our approach more flexible, and generalizable to weighted graphs, we used two variants of the above perturbation framework: (1) the first scenario ($\delta_{w}=0$) is exactly the one described above, i.e., it ignores edge weights during the perturbation phase. Edges can only be completely removed or created from scratch; (2) in the second scenario ($\delta_{w}=1$), the perturbation procedure accounts for edge weights. More precisely, a given edge $(u,v)$ can be considered for addition even if it already exists, and can remain in the graph even if it was selected for deletion. In such cases, we simply increment (respectively, decrement) the weight of $(u,v)$ by one standard deviation of all edge weights. Since edges can be selected multiple times, any edge whose weight becomes negative is removed from the graph.
In both scenarios, whenever an edge is created, we sample its weight at random from the weights of the edges incident on its endpoints.

\noindent \textbf{Random graph models}. Plugging the Erd\H{o}s-R\'{e}nyi (\textbf{ER}) random graph model \cite{erdos} into our framework returns the \textit{uniform perturbation model}. A node is randomly drawn with replacement from $V$ with probability $1/n$. On the other hand, using the Chung-Lu (\textbf{CL}) random graph model \cite{chung2002average} gives the \textit{degree assortative perturbation model}. A node is randomly drawn with replacement from $V$ with probability proportional to its weighted degree. In that case, edges are more likely to be created/incremented and deleted/decremented between hubs. For both $\mathbb{G}=ER$ and $\mathbb{G}=CL$, self-edges are disregarded. That is, if we select two nodes $u$ and $v$ such that $u=v$, we discard the pair and select two other nodes.

\subsection{Mine}\label{sub:mine}
Vertex scores can be returned by any node scoring function. However, since perturb and combine strategies in Machine Learning are most effective when used with \textit{unstable} base learners, we experimented with $k$-core and weighted $k$-core ($cu$ and $cw$ in what follows), as these algorithms are both unstable and highly effective at locating good spreaders. Since its rankings are known to be relatively stable to link-based perturbations\cite{ipsen2006mathematical,ng2001link}, we also included the weighted PageRank algorithm \cite{page1999pagerank} in our experiments, for comparison purposes ($pr$ in what follows).

\subsection{Combine}\label{sub:combine}
Following common practice in Machine Learning (e.g., bagging regression trees), we compute the P\&C score of a node as the average of its scores over the $M$ perturbed graphs.

\section{Experiments}\label{sec:exps}
We tested our P\&C strategy on large social networks and small word co-occurrence networks.

\subsection{Social Networks}\label{sub:social}
We experimented on 3 well-known, publicly available large scale social networks \cite{snapnets} (see Table \ref{datasets}). \textbf{Email-Enron} is an email communication network \cite{klimt2004enron} where edges indicate email exchange. \textbf{Epinions} is a social network where edges indicate who trusts whom. In the \textbf{WikiVote} network, edges represent administrator election votes. Since these networks are unweighted, we assigned as edge weights the maximum number of connections of their endpoints.

\begin{table}[h]
\captionsetup{size=small}
\setlength{\belowcaptionskip}{-4pt}
 \caption[]{Statistics of the real-world social networks used in this study. $d$ and $\tau$ denote diameter and epidemic threshold. \label{datasets}}	
 \centering
\scalebox{0.77}{
 	\begin{tabular}{lccccccc}
 		\toprule
 		$G(V,E)$  & \textbf{$|V|$} & \textbf{$|E|$}& $d$  & $cu_{max}$  & $cw_{max}$ & $pr_{max}$ &$\tau \times 10^2$\\
 		\midrule
 		\textsc{Email-Enron}	& $33,696$    & $180,811$ & $11$ & $43$& $18,931$& $14.95$ &  $8.4$  \\
 		\textsc{Epinions}	& $75,877$    & $405,739$  & $14$&  $67$ & $37,343$& $3.06$ & $5.4$ \\
 		\textsc{Wiki-Vote}	& $7,066$    & $100,736$ & $7$&  $53$ &  $15,756$ & $4.37$& $7.2$ \\
 		\hline
 	\end{tabular}
 }
\end{table}
 
\vspace{-0.35cm}
 
\noindent To measure the spreading influence of a node, we simulated diffusion processes using the famous Susceptible-Infected-Recovered (SIR) epidemic model \cite{kermack1927contribution}. It is a discrete time model which assumes that at every step, each node falls into one of the following mutually exclusive categories:

\noindent $\bullet$ Susceptible (S): not yet infected, thus being able to get infected  with probability $\beta$ (infection rate),

\noindent $\bullet$ Infected (I): capable of contaminating susceptible neighbors,

\noindent $\bullet$ Recovered (R): after being infected, a node may recover with probability $\gamma$ (recovery rate). It is then considered as immune and cannot transmit infection anymore. The process iterates until no new node gets infected for two consecutive steps.

\vspace{-0.45cm}

\begin{figure}[h]
  \begin{center}
  \captionsetup{size=small}
  	\tikzset{
  		int/.style={circle, draw, fill=yellow!20, minimum size=2.5em}
  	}
  	\scalebox{0.6}{
  	\begin{tikzpicture} [node distance=2.5cm,auto,>=latex',every node/.append style={align=center}, thick]	
  	\node [int] (a)   {$S$};
  	\node [int](c) [right of=a] {$I$};
  	\node [int] (e) [right of=c] {$R$};
  	
  	\draw[->, line width=1pt] (a) edge node {$\beta$} (c); 
  	\draw[->, line width=1pt] (c) edge node {$\gamma$} (e);
  	\draw[->, line width=1pt] (a) edge [loop above] node {$1-\beta$} (a);
  	\draw[->, line width=1pt] (c) edge [loop above] node {$1-\gamma$} (c);
  	\end{tikzpicture}
  	}
  	\setlength{\belowcaptionskip}{-11pt}
  	\caption{SIR epidemic model}
  \end{center}
\end{figure}
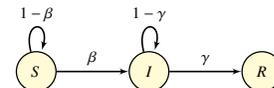 

\noindent Following \cite{kitsak2010identification,malliaros2016locating}, we started $N_e$ epidemics from each node in the trigger population to account for the stochastic nature of the SIR model. The results for each node were then averaged over the $N_e$ runs and then again averaged over the trigger population to get a final performance score for a given vertex scoring function.

For unweighted and weighted $k$-core, the trigger population was the maximal $k$-core subgraph. Since we used averaging as our combination strategy, we rounded up the P\&C scores to the nearest integer before extracting the main cores. For PageRank, the trigger population was the 100 nodes with highest scores. We set $N_e=100$, the infection rate $\beta$ close to the epidemic threshold of the network $\tau=\frac{1}{\lambda_1}$, where $\lambda_1$ was the largest eigenvalue of the adjacency matrix \cite{chakrabarti2008epidemic}, and the recovery rate $\gamma$ to $0.8$, as in \cite{kitsak2010identification}. We grid searched the following P\&C parameters: $\varepsilon_a:\{0, 0.05, 0.1, 0.2\}$, $\varepsilon_d:\{0, 0.05, 0.1, 0.2\}$, $M:\{16,64\}$, $\mathbb{G}:\{ER,$ $CL\}$, and $\delta_{w}:\{0,1 \}$. Excluding the $\varepsilon_a = \varepsilon_d = 0$ cases, this made for 120 combinations. The optimal parameter values were selected as the ones returning the greatest total number of nodes infected during the epidemic.

\textbf{Results}. Table \ref{table:res_social} compares the average severity of the epidemic when triggered from the top nodes in terms of their scores in the original networks to that when started from the top nodes in terms of P\&C scores, for the best parameters shown in Table \ref{table:best_social}. Everywhere, using the P\&C scores leads to a more severe epidemic.

\begin{table}[h]
\setlength\tabcolsep{3pt}
\renewcommand{\arraystretch}{0.65}
\scriptsize
\setlength{\belowcaptionskip}{-4pt}
\setlength{\abovecaptionskip}{-4pt}
\captionsetup{size=small}
\caption{Epidemic severity for unweighted $k$-core, weighted $k$-core, and PageRank (top to bottom). +\% denotes percent improvement.}\label{table:res_social}
	\begin{center}	
	    	\scalebox{1}{
	\begin{tabular}{lrccccccr}
        	&  &\multicolumn{5}{c}{Time Step} & \\
				Network & Scores & 2  & 4  & 6 & 8  & 10 & Total & +\% \\
				\hline
				\textsc{Enron}&\textbf{P\&C} & \textbf{16} &  \textbf{89} &  \textbf{300} &  \textbf{419}  & \textbf{269}  & \textbf{2,538} & 3.76\\
				 &original & 14 &  77 &  269 &  401  & 275  & 2,446 \\
				\textsc{Epinions}&\textbf{P\&C} & \textbf{8} &  \textbf{34} &  \textbf{110} &  \textbf{245}  & \textbf{317}  & \textbf{2,436} & 4.35\\
				&original & 7 &  30 &  100 &  224  & 301  & 2,330 \\
				\textsc{WikiVote}&\textbf{P\&C} & \textbf{3} &  \textbf{8} &  \textbf{17} &  \textbf{29}  & \textbf{40}  & \textbf{490} & 3.47\\
				&original & 3 &  8 &  16 &  28  & 37  & 473 \\
				\hline
			\end{tabular}
			}
			
		    	\scalebox{1}{	\begin{tabular}{lrllllllr}
				\textsc{Enron}&\textbf{P\&C} & \textbf{26} &  \textbf{141} &  \textbf{407} &  \textbf{445}  & \textbf{226}  & \textbf{2,724} & 3.52\\
				&original & 20 &  110 &  345 &  433  & 253  & 2,628 \\
				\textsc{Epinions}&\textbf{P\&C} & \textbf{11} &  \textbf{46} &  \textbf{146} &  \textbf{302}  & \textbf{353}  & \textbf{2,689} & 2.42 \\
				&original & 11 &  42 &  135 &  286  & 345  & 2,624 \\
				\textsc{WikiVote}&\textbf{P\&C} & \textbf{5} &  \textbf{12} &  \textbf{24} &  \textbf{39}  & \textbf{50}  & \textbf{612} & 19.3\\
				&original & 4 &  9 &  18 &  31  & 42  & 513 \\
				\hline
			\end{tabular}
			}
			
		    	\scalebox{1}{	\begin{tabular}{lrccccccr}
				\textsc{Enron}&\textbf{P\&C} & \textbf{16} &  \textbf{86} &  \textbf{278} &  \textbf{389}  & \textbf{266}  & \textbf{2,454} & 4.93 \\
				& original & 15 &  80 &  259 &  366  & 255  & 2,333 \\
				\textsc{Epinions}&\textbf{P\&C} & \textbf{11} &  \textbf{42} &  \textbf{132} &  \textbf{276}  & \textbf{336}  & \textbf{2,598} & 2.04\\
				&original & 11 &  41 &  127 &  267  & 326  & 2,545 \\
				\textsc{WikiVote}&\textbf{P\&C} & \textbf{5} &  \textbf{11} &  \textbf{22} &  \textbf{38}  & \textbf{49}  & \textbf{596} & 2.35 \\
				&original& 5 &  11 &  22 &  36  & 48  & 582 \\
				\hline
			\end{tabular}
			}
	\end{center}
\end{table}

\vspace{-0.5cm}

\begin{figure}[h]
    \centering
\setlength\tabcolsep{3pt}
\renewcommand{\arraystretch}{0.42}
\scriptsize
    \setlength{\belowcaptionskip}{-12pt}
    \captionsetup{size=small}
	\includegraphics[width=0.95\columnwidth]{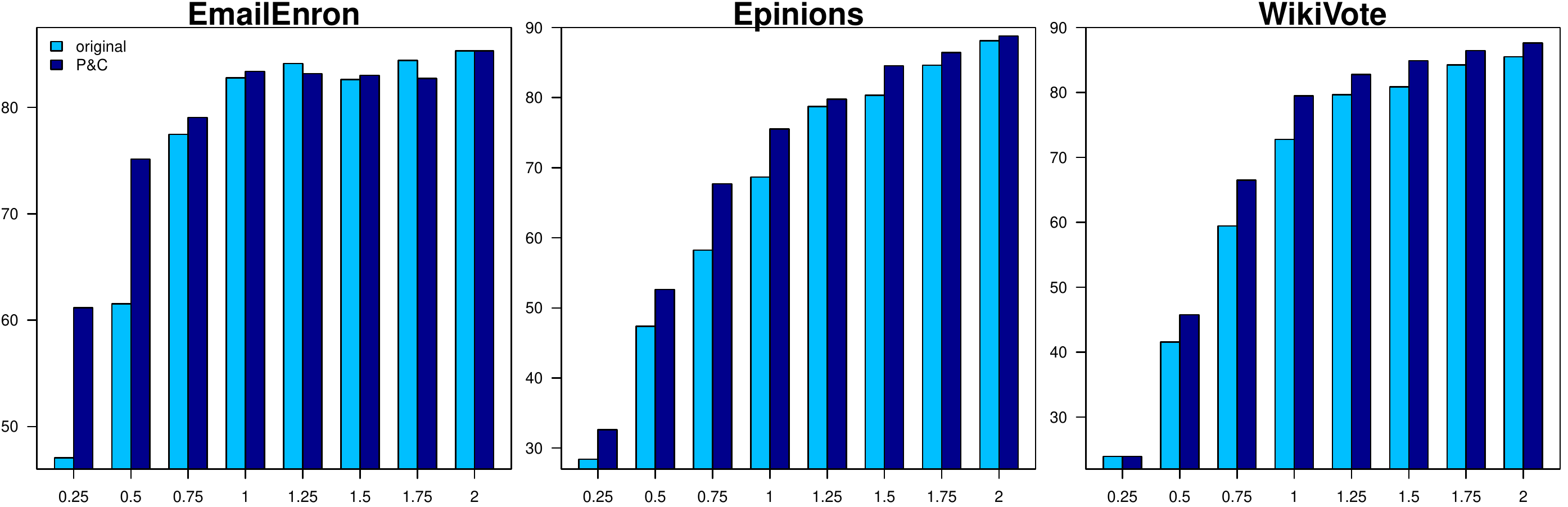}
	\scalebox{1}{
	\begin{tabular}{rcccccccc}
	$p$ & 0.25 & 0.5 & 0.75 & 1 & 1.25 & 1.5 & 1.75 & 2 \\ 
\hline
ENRON & 85 & 169 & 253 & 337 & 422 & 506 & 590 & 674 \\ 
  WIKIVOTE & 71 & 142 & 212 & 283 & 354 & 424 & 495 & 566 \\ 
  EPINIONS & 95 & 190 & 285 & 380 & 475 & 570 & 664 & 759 \\ 
  \hline
\end{tabular}
}
	\caption{Fraction of the p\% best spreaders (y axis) contained in the top p\% nodes (x axis) in terms of P\&C and original scores, for $cu$. The table shows the number of best spreaders for each $p$. \label{fig:social_barplots}}
\end{figure}

Moreover, the differences are substantial, comparable with the improvements reported in previous research, e.g., $k$-truss over $k$-core \cite{malliaros2016locating}. We can also remark that even though the parameters were tuned to maximize the total number of nodes infected during the entire epidemic, nodes with high P\&C scores are better spreaders even in the early stages of the diffusion process. Finally, as shown in Fig. \ref{fig:social_barplots} for unweighted $k$-core, the rankings provided by P\&C are of better quality than the rankings provided by the original scores, especially for the very top nodes. E.g., on Enron, the top 0.5\% (169) P\&C nodes contain $75.15\%$ of the 0.5\% best spreaders (as returned by SIR), compared to only $61.54\%$ when using the original scores. This means that P\&C places \textit{more of the most influential spreaders} at the very top of the ranking than the original scores. This is a very valuable property, especially in practice, when budget constraints often only allows selecting a limited number of nodes (e.g., in growth hacking and viral marketing).

\vspace{-0.25cm}

\subsection{Word co-occurrence networks}\label{sub:nlp}
\begin{figure}[H]
    \centering
    \captionsetup{size=small}
	\includegraphics[width=.75\textwidth]{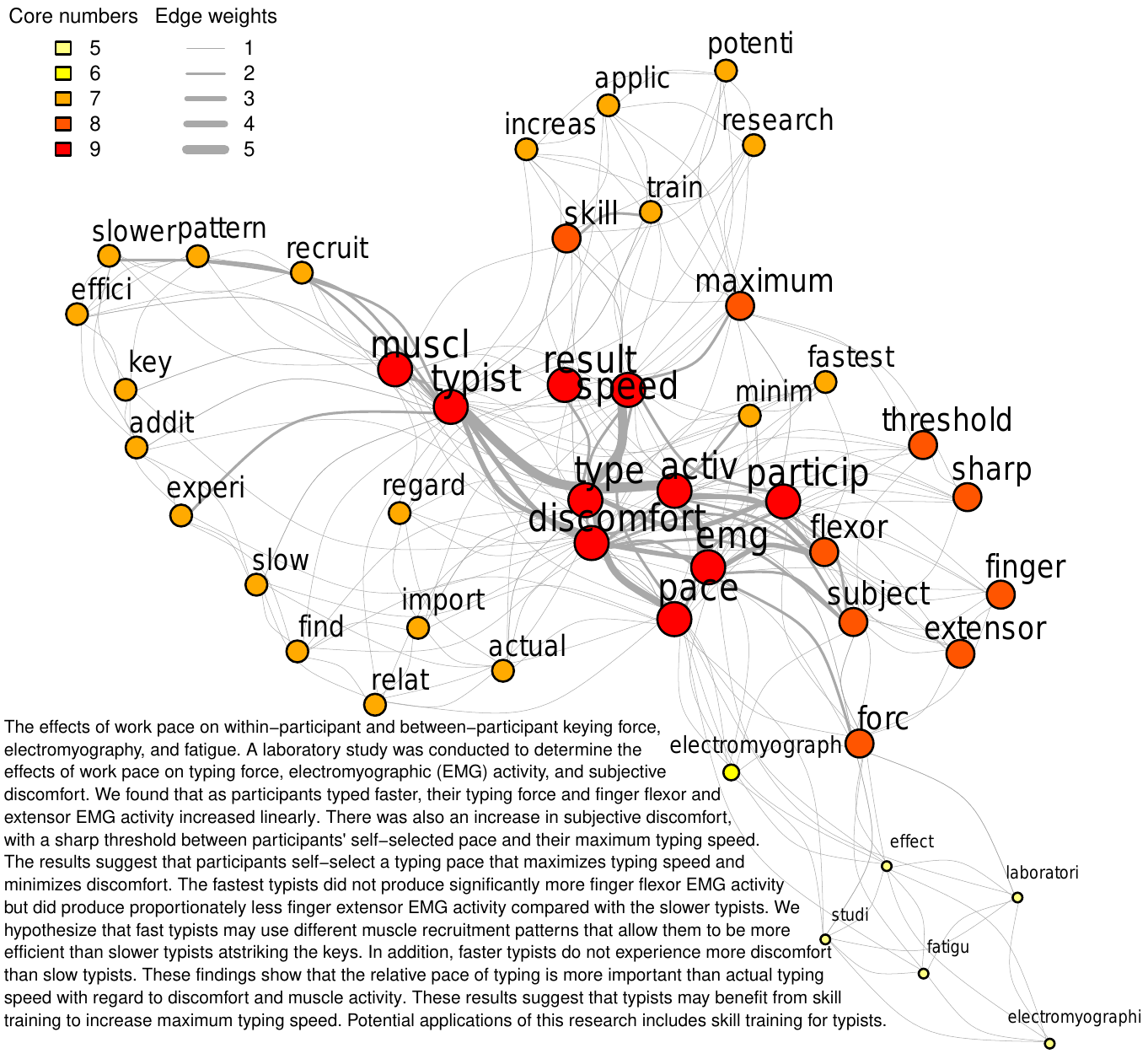}
	\caption{Word co-occurrence network representation of document 1478 of the Hulth 2003 dataset. Only nouns and adjective are kept (and then stemmed). $W=5$. The human keywords (stemmed) for this document are \textit{work, pace, effect, emg, activ, subject, discomfort, finger, flexor, type, speed, typist, muscl, recruit, pattern, kei, forc, skill, train}.
	\label{fig:gow}}
\end{figure}

It has been suggested that the keywords of a document are influential nodes within the word co-occurrence network of that document \cite{tixier2016graph}. Therefore, in this section, we test whether applying P\&C to graphs of words improves keyword extraction performance.
We define a word co-occurrence network as in \cite{mihalcea2004textrank,tixier2016gowvis} (see Fig. \ref{fig:gow} in App. D). Each unique term is a node, and two nodes are linked by an edge if they co-occur within a fixed-size sliding window. Edge weights indicate co-occurrence counts.

We experimented on the well-known Hulth2003 dataset \cite{hulth2003improved}\footnote{\label{note1}\url{https://github.com/snkim/AutomaticKeyphraseExtraction}}, which contains abstracts from the Inspec research article database. We considered the standard validation set of 500 documents and used the uncontrolled keywords assigned by human annotators as ground truth. The mean document size is 120 words and on average, 21 keywords (unigrams) are available for each document. Note that reaching perfect recall is impossible on this dataset.

Following \cite{mihalcea2004textrank}, we pre-processed each document with part-of-speech (POS) tagging and retained only nouns and adjectives. Finally, we Porter stemmed words. We then built a word co-occurrence network for each document with a window of size 5. Human keywords were stemmed too, but they were not POS-filtered, i.e., they contained verbs. 
The average number of nodes, edges, and diameter of the networks were respectively $32$, $155$, and $3.6$. 
For unweighted and weighted $k$-core, we retained as keywords the words belonging to the maximal $k$-core subgraph. Again, since our combination strategy was averaging, we rounded up the P\&C scores to the nearest integer before extracting the main cores. For PageRank, we extracted the top 33\% nodes as keywords. The following P\&C parameters were grid searched: $\varepsilon_a:\{0, 0.1, 0.2, 0.3\}$, $\varepsilon_d:\{0, 0.1, 0.2, 0.3\}$, $M:\{8,32,96\}$, $\mathbb{G}:\{ER,$ $CL\}$ and $\delta_{w}:\{0,1\}$. Excluding the $\varepsilon_a = \varepsilon_d = 0$ cases, this made for 180 combinations.

\noindent \textbf{Results}.
Performance is reported in terms of the usual macro-averaged precision, recall, and F1-score in Table \ref{table:results_nlp}. The results for P\&C are that obtained with the best parameter combination for each scoring function, reported in Table \ref{table:best_nlp}.

\begin{table}[h]
\captionsetup{size=small}
\setlength{\belowcaptionskip}{-4pt}
\setlength{\abovecaptionskip}{-4pt}
\caption{Upper part: keyword extraction results. +\% denotes percent improvement. Lower part: best results reported in SOTA unsupervised graph-based keyword extraction papers.}\label{table:results_nlp}
	\begin{center}					
\resizebox{0.85\columnwidth}{!}{
			\begin{tabular}{lccccl}
				$s$ & scores & precision  & recall  & F1-score & +\% \\
				\hline
				$cu$&\textbf{P\&C} & \textbf{52.09} &  \textbf{51.25} &  \textbf{54.88} & 5.70\\
				&original & 48.76 &  46.90 &  51.75 \\
				$cw$&\textbf{P\&C} & \textbf{50.53} &  \textbf{48.54} &  \textbf{52.50} & 7.45 \\
				&original & 48.07 &  46.81 &  48.86 \\
				$pr$&\textbf{P\&C} & \textbf{45.53} &  \textbf{42.73} &  \textbf{46.75} & 2.33\\
				&original& 45.21 &  41.89 &  45.66 \\
				\hline
				\hline
				& [Tixier16]\cite{tixier2016graph} & 48.79 & 72.78 & 56.00 & \\
				& [Rousseau15]\cite{rousseau2015main} & 61.24 & 50.32 & 51.92 \\
				&  [Mihalcea04] \cite{mihalcea2004textrank} & 51.95 & 54.99 & 50.40 & \\
			\end{tabular}
		}
		
	\end{center}
\end{table}

\vspace{-0.42cm}

\noindent Like on social networks, using the P\&C scores in place of the original scores improves performance for every algorithm, with large improvements ranging from 1.09 to 3.64 in F1-score. Even though looking at improvements is sufficient to show that our P\&C strategy is effective, it is to be noted that our scores are on par with or exceed the state-of-the-art (SOTA) in the field of unsupervised graph-based keyword extraction, as shown in the lower part of Table \ref{table:results_nlp}.

\section{Theoretical analysis}\label{sec:theory}
We now try to understand why P\&C for networks is effective from a theoretical perspective. The notation used in this section is summarized in Table \ref{tab:notation}.

\noindent \textbf{Underlying graph}.
Let us assume the existence of a true but unavailable underlying graph $G^*$, of which the available graph $G$ is a snapshot, or sample, so that $G$ features the same nodes as $G^*$ but has a slightly different edge set. This is analogous to the traditional assumption made in statistics that \textit{a given dataset represents a sample of a true but unknown distribution}. Since $G^*$ is unavailable, we have to find a way to emulate sampling from $G^*$ by using only $G$. The solution we opt for is to perturb $G$.

\begin{table}[H]
\captionsetup{size=small}
\setlength{\belowcaptionskip}{-4pt}
\setlength{\abovecaptionskip}{-1pt}
\caption{\label{tab:notation}Notation summary.}
\centering
\resizebox{0.95\columnwidth}{!}{
\begin{tabular}{|ll|}
\hline
    $G(V,E)$ & undirected weighted graph, $n=|V|$ \\ 
	$v_i \in V$ & the $i$-th vertex of $G$, $i \in \{1,\ldots,n\}$ \\
	$G^*$ & true (unknown) graph underlying $G$ \\
	$G \sim G^*$ & available graph (sample, snapshot of $G^*$) \\
	$\Gapprox_m \sim G$ & $m$-th perturbed version of $G$ \\
	$\{\Gapprox\}$ & set of $M$ perturbed graphs $\{\Gapprox\}_{m=1}^M$ \\
	$s : V \mapsto \mathbb{R}^{|V|}$ & vertex scoring function, e.g., $k$-core algorithm  \\
	$s_{G}(i)$ & score of vertex $v_i$ in $G$ \\
    $R = \{l(1),\ldots,l(n)\}$ & true ranking on $G$ \\
    $\Rest$ & ranking when $s$ is applied on $G$, estimate of $R$ \\ 
    $\Rest_m$ & ranking when $s$ is applied on $\Gapprox_m$; estimate of $R$ \\ 
    $\text{MET}$ & goodness of fit of $\Rest$ w.r.t. $R$ \\  
\hline
\end{tabular}
}
\end{table}

\vspace{-0.2cm}

\noindent \textbf{True ranking}. 
Let us also assume the existence of a true ranking $R = \{l(1),\ldots,l(n)\}$ of the nodes $\big\{v_{1}, \dots, v_{n}\big\}$ of $G$ in terms of their spreading capability, that associates each node $v_i$ with one of $K$ labels $l(i)$. $v_i$ is ranked before $v_j$ if $l(i) > l(j)$. $K \leq n$ as some nodes may be equally good spreaders. This true ranking can be a given, or can be computed, for instance with the SIR model.

\noindent \textbf{Objective}. Let $s: V \mapsto \mathbb{R}^{|V|}$ be a vertex scoring function, i.e., a function that associates each node of $G$ with a real number, and let $\Rest$ be the ranking induced by $s$ on the nodes of $G$. $\Rest$ can be seen as an \textit{estimate} of the true ranking $R$, and $s$ as an \textit{estimator}. Let us also assume that the quality of the estimate provided by $s$ (goodness of fit) is measured by a metric $\mathrm{MET}$ accepting $\Rest$ and $R$ as input and taking values in $\big[0,1\big]$. The objective of $s$ is to maximize $\mathrm{MET}$. $\mathrm{MET}$ is a random variable (RV) as $\Rest$ is a RV.

\noindent \textbf{Perturbation as sampling}. In each of the $M$ edge-perturbed version $\tilde{G}_m$ of $G$, the individual node scores, and by extension the rankings $\Rest_m$, randomly vary, as our perturbation strategy is stochastic. We can thus consider the $\Rest_m$ to be RVs. Moreover, since the $\tilde{G}_m$ are generated independently, the $\Rest_m$ are independent. Therefore, perturbing $G$ is akin to \textit{sampling independent realizations from the true underlying graph $G^*$}. Scoring nodes based on different configurations of the original network is thus akin to estimating the true scores of the nodes based on more evidence (see App. B).

\noindent \textbf{Definitions: bias and variance of a vertex scoring function}.
Our goal is to study the impact of P\&C on the goodness of fit of $s$. In regression, the error is traditionally decomposed into bias and variance terms. We adopt this framework and define in what follows the bias and variance of $s$.

\noindent In the regression setting, $y = f(x) + \epsilon$, $\sigma^2=\mathrm{var}\big[\epsilon\big]$, $\hat{f}$ is an estimator of $f$, and we have the following well-known breakdown of expected squared error of the estimation into (squared) bias, variance, and irreducible error terms:

\vspace{-0.5cm}

\begin{align}
\mathbb{E}\big[(y-\hat{f}(x))^2\big] &= \mathrm{bias}\big[\hat{f}(x)\big]^2 + \mathrm{var}\big[\hat{f}(x)\big] + \sigma^2 \\
\mathrm{bias}\big[\hat{f}(x)\big] &=\mathbb{E}\big[\hat{f}(x)-f(x)\big] \\
\mathrm{var}\big[\hat{f}(x)\big]&=\mathbb{E}\big[\big(\hat{f}(x)-\mathbb{E}\big[\hat{f}(x)\big]\big)^2\big]
\end{align}

\noindent The expectation is computed for different samples drawn from the same underlying distribution. By analogy, in our setting, we can define the bias and variance of $s$ as:

\vspace{-0.25cm}

\begin{equation}
\label{eq:bias}
\mathrm{bias}\big[s\big]=\mathbb{E}\big[1-\mathrm{MET}\big] = 1- \mathbb{E}\big[\mathrm{MET}\big]
\end{equation}
\begin{equation}
\label{eq:var}
\mathrm{var}\big[s\big] = \mathbb{E}\big[\big(\mathrm{MET} - \mathbb{E}\big[\mathrm{MET}\big]\big)^2\big]
\end{equation}

\vspace{-0.25cm}

\noindent The bias captures, on average, how close the estimated ranking $\Rest$ provided by $s$ is to the true ranking $R$ (for which $\mathrm{MET}$ is equal to 1), while the variance measures the instability of $\Rest$ (variability around its mean). The expectation is to be understood as computed over a set of observations of $G^*$. Since for all RVs $X,Y$ and $k \in \mathbb{R}$, $\mathbb{E}\big[X+Y\big] = \mathbb{E}\big[X\big] + \mathbb{E}\big[Y\big]$, $\mathbb{E}\big[k\big] = k$, and $\mathbb{E}\big[kX\big] = k\mathbb{E}\big[X\big]$, developing Eq. \ref{eq:var} gives:

\vspace{-0.5cm}

\begin{align}
\mathrm{var}\big[s\big] & = \mathbb{E}\big[\mathrm{MET}^2 - 2\mathbb{E}\big[\mathrm{MET}\big]\mathrm{MET} + \mathbb{E}^2\big[\mathrm{MET}\big] \big] \\ 
 & = \mathbb{E}\big[\mathrm{MET}^2\big] - 2 \mathbb{E}^2\big[\mathrm{MET}\big] + \mathbb{E}^2\big[\mathrm{MET}\big]
\end{align}

\noindent Summing the squared bias and variance terms thus gives:

\vspace{-0.5cm}

\begin{align}
\mathrm{bias}^2\big[s\big] + \mathrm{var}\big[s\big]
& = 1 -2\mathbb{E}\big[\mathrm{MET}\big] + \mathbb{E}\big[\mathrm{MET}^2\big] \label{eq:squared_error} \\
& = \mathbb{E}\big[\big(\mathrm{MET}-1\big)^2\big]
\end{align}

\noindent which can be interpreted as the expectation of the squared error, like in the case of regression. The proof of Eq. \ref{eq:squared_error} is given in App. A.\\

\vspace{-0.3cm}

\noindent \underline{\textbf{Theorem: P\&C reduces error}}.\\
\noindent \textbf{Proof}.
Recall that the P\&C score $s_{\mathrm{pc}}$ of node $v_i$ is defined as the average of the scores its gets in each of the $M$ perturbed graphs $\{\Gapprox\} = \{\tilde{G}_m\}_{m=1}^M$ generated from $G$:

\vspace{-0.15cm}

\begin{equation}
\label{eq:s_pc}
s_{\mathrm{pc}}(v_i) = \frac{1}{M}\sum_{m=1}^{M} s_{\Gapprox_m}(v_i)
\end{equation}

\noindent By definition, we can also write:

\vspace{-0.15cm}

\begin{equation}
\Rest_{{\mathrm{pc}}} = \mathbb{E}_{\big\{\tilde{G}\big\}}\big[\big\{\Rest\big\}\big]
\end{equation}

\noindent where $\big\{\Rest\big\} = \{\Rest_m\}_{m=1}^M$. That is, the P\&C estimate $\Rest_{\mathrm{pc}}$ of the true ranking $R$ is the average of the estimates $\Rest_m$ over the $M$ perturbed graphs. Similarly, the goodness of fit of the P\&C ranking can be written: 

\vspace{-0.15cm}

\begin{equation}\label{eq:metric_pc}
\mathrm{MET}_{\mathrm{pc}} = \mathbb{E}_{\big\{\tilde{G}\big\}}\big[\mathrm{MET}\big]
\end{equation}

\noindent where $\mathrm{MET}\big( \big\{\Rest\big\}, R\big)$ is simply written $\mathrm{MET}$ for readability. Thus, evaluating Eq. \ref{eq:squared_error} over $\big\{\tilde{G}\big\}$, and using Eq. \ref{eq:metric_pc} above:

\vspace{-0.3cm}

\begin{align}
\mathbb{E}_{\big\{\tilde{G}\big\}}\big[\big(\mathrm{MET}-1\big)^2\big] & = 1 - 2\mathbb{E}_{\big\{\tilde{G}\big\}}\big[\mathrm{MET}\big] + \mathbb{E}_{\big\{\tilde{G}\big\}}\big[\mathrm{MET}^2\big] \\
&  = 1 - 2\mathrm{MET}_{\mathrm{pc}} + \mathbb{E}_{\big\{\tilde{G}\big\}}\big[\mathrm{MET}^2\big]
\end{align}

\noindent Plus, since for all RV $X$ and $k \in \mathbb{R}$, $\mathbb{E}^2\big[X\big] \geq \mathbb{E}\big[X^2\big]$ and $\mathbb{E}\big[k\big] = k$, using again Eq. \ref{eq:metric_pc}, and since $\mathbb{E}$ is monotone:

\vspace{-0.3cm}

\begin{align}
\mathbb{E}_{\big\{\tilde{G}\big\}}\big[\big(\mathrm{MET}-1\big)^2\big] & \geq 1 - 2\mathrm{MET}_{\mathrm{pc}} + \mathbb{E}^{2}_{\big\{\tilde{G}\big\}}\big[\mathrm{MET}\big] \\
& \geq \big(1 - \mathrm{MET}_{\mathrm{pc}}\big)^2 \\
\mathbb{E}_{\big\{\tilde{G}\big\}}\big[\big(\mathrm{MET}-1\big)^2\big] & \geq \mathbb{E}_{\big\{\tilde{G}\big\}}\big[\big(1 - \mathrm{MET}_{\mathrm{pc}}\big)^2\big] \label{eq:inequality}
\end{align}

\noindent Inequality \ref{eq:inequality} shows that \textbf{\textit{the mean squared error of P\&C (RHS) is always lower than or equal to the original mean squared error (LHS)}}, which is an important result. Improvement can come from reducing bias and/or variance. \QEDB

\noindent \textbf{Sample bias and variance}. To understand how Inequality \ref{eq:inequality} holds in practice, we randomly selected 16 Hulth2003 word networks, and generated 50 perturbed version of each. As previously explained, this can be considered as drawing 50 independent realizations from the underlying graph that generated each network. With unweighted $k$-core, we then scored the nodes of each graph in the sample with and without using our P\&C strategy, and computed the goodness of fit of each ranking, using the Normalized Discounted Cumulative Gain (NDCG) \cite{jarvelin2002cumulated} as the metric $\mathrm{MET}$. 

The NDCG is a standard metric for assessing ranking quality in Information Retrieval (IR). In our case, the more of the most influential spreaders are placed on top of $\Rest$, the better the NDCG. More precisely, NDCG is computed as $\nicefrac{\mathrm{DCG}}{\mathrm{IDCG}}$, where DCG is the Discounted Cumulative Gain computed on $\Rest$ and IDCG is the ideal DCG computed on $R$. NDCG is maximal and equal to 1 if $\Rest$ matches $R$ exactly. Generally in IR, the DCG is computed over a shortlist of the best results, but we can assume without loss of generality that it is computed over the full list of $n$ nodes:

\vspace{-0.25cm}

\begin{equation}
\mathrm{DCG} = \sum_{i=1}^{n} \frac{2^{rel_i} - 1}{\mathrm{log}_2(i+1)}
\end{equation}

\vspace{-0.1cm}

\noindent where $i$ designates the rank of node $v_i$ in the list. We used as the relevance score $rel_i$ of $v_i$ its SIR influence, i.e., the average number of nodes infected at the end of multiple epidemics triggered from it. We finally computed bias and variance from the set of 50 NDCGs by using Eq. \ref{eq:bias} and Eq. \ref{eq:var}. We repeated the same procedure for the WIKIVOTE network. 

Results are shown in Table \ref{table:sample_gow}. As can be seen for word networks, P\&C  reduces both bias and variance, although its major contribution appears to lie in the consistent reduction of bias.
The average of the averages of the NDCGs is 0.58 for the rankings obtained with the original scores and 0.73 for P\&C, which means that the rankings returned by the P\&C scores fit the true rankings much better.
On WIKIVOTE, P\&C reduces bias, but not variance, which is initially very low. The P\&C NDCG is 0.203 while the original is 0.189, indicating again that P\&C returns better rankings.

\begin{table}[]
\setlength\tabcolsep{3pt}
\renewcommand{\arraystretch}{0.8}
\scriptsize
\captionsetup{size=small}
\setlength{\belowcaptionskip}{-4pt}
\setlength{\abovecaptionskip}{-1pt}
\caption{Left and middle tables: sample bias ($\times10^2$) and variance ($\times10^3$) for 16 randomly selected word networks. Right: sample bias ($\times10^2$) and variance ($\times10^5$) for WIKIVOTE. \textbf{Lower} is better. \label{table:sample_gow}}
\centering
\scalebox{1.1}{
\begin{tabular}{rccc}
& original & P\&C \\ 
  \hline
bias & 37.06 & \textbf{24.92} \\ 
  var & 4.78 & \textbf{1.17} \\ \hline
  bias & 41.92 & \textbf{16.66} \\ 
  var & 12.83 & \textbf{4.72} \\ \hline
  bias & 48.44 & \textbf{35.64} \\ 
  var & 6.02 & \textbf{5.43} \\ \hline
  bias & 42.86 & \textbf{30.23} \\ 
  var & 0.08 & 4.44  \\ \hline
  bias & 45.16 & \textbf{31.23} \\ 
  var & 0.03 & 4.09  \\ \hline
  bias & 64.56 & \textbf{38.97} \\ 
  var & 1.87 & 4.04  \\ \hline
  bias & 24.75 & \textbf{14.05} \\ 
  var & 0.58 & 2.52  \\ \hline
  bias & 31.72 & \textbf{19.00} \\ 
  var & 0.58 & 6.78 \\
  \hline
\end{tabular}
}
\quad
\scalebox{1.1}{
\begin{tabular}{rccc}
& original & P\&C \\
  \hline
  bias & 51.59 & \textbf{34.97}\\
  var & 0.66 & 1.16 \\ \hline
  bias & 39.16 & \textbf{27.22}\\ 
  var & 0.04 & 1.79 \\ \hline
  bias & 66.43 & \textbf{54.66}\\ 
  var & 0.09 & 0.76 \\ \hline
  bias & 46.97 & \textbf{21.98}\\ 
  var & 1.54 & 2.71 \\ \hline
  bias & 26.40 & \textbf{20.75}\\ 
  var & 0.04 & 0.49 \\ \hline
  bias & 37.29 & \textbf{24.40}\\ 
  var & 0.27 & 2.17 \\ \hline
  bias & 33.76 & \textbf{18.68}\\ 
  var & 0.18 & 2.50 \\ \hline
  bias & 41.61 & \textbf{24.12}\\ 
  var & 6.85 & \textbf{6.60}\\
  \hline
  \end{tabular}
  }
  \scalebox{1.1}{
  \begin{tabular}{lrrr}
& original & P\&C \\ 
  \hline
bias & 81.08 & \textbf{79.68}\\ 
  var & 0.00 & 0.64\\
  \hline
   & & \\
   & & \\
   & & \\
   & & \\
   & & \\
   & & \\
   & & \\
   & & \\
   & & \\
   & & \\
   & & \\
   & & \\
   & & \\
   & & \\
   & & \\
\end{tabular}
}
\end{table}

\noindent \textbf{P\&C for networks differs from bagging}. Our P\&C strategy reducing mainly bias rather than variance suggests that it differs from bootstrap aggregation (bagging). Indeed, bagging can increase variance when it fails \cite{BaggingEqualizesInfluence}, but it is widely accepted that it cannot significantly reduce bias, as illustrated by the fact that it does not work well with stable algorithms such as nearest-neighbors approaches \cite{breiman1996bagging}. Another obvious difference is that a bootstrap sample has always the same size as the original dataset, whereas in our case, a perturbed graph has as many edges as the original graph only when $\varepsilon_a = \varepsilon_d$.

\noindent To strictly emulate bagging, we would need to adopt a different framework in which we would draw \textit{edges} from a true underlying distribution rather than graphs. The perturbation step would only consist in sampling edges with replacement from the original set of edges. Some edges would be selected more than once, while some edges would not be selected at all. In the final perturbed network, 63.2\% of unique edges would carry over from the original network, but no new edge would be present. This would remove the need for the $\varepsilon_a$ and $\varepsilon_d$ parameters, at the cost of losing flexibility. The improvement brought by P\&C would be obtained mainly by reducing variance as:

\vspace{-0.35cm}

\begin{equation}\label{eq:varvar}
\begin{aligned}
\mathrm{var}\big[\Rest_{\mathrm{pc}}\big] & = \mathrm{var}\Bigg[\frac{1}{M}\sum_{m=1}^{M} \Rest_{m}\Bigg] \\
& = \frac{1}{M^2} \sum_{m=1}^{M} \mathrm{var}\big[ \Rest_{m} \big] \\
\end{aligned}
\end{equation}

\vspace{-0.15cm}

\noindent The extent to which error would be reduced would thus only depend on the amount of uncorrelation among the rankings $\Rest_m$, as correlation adds positive covariance terms to the RHS of Eq. \ref{eq:varvar}, i.e., increases variance. In other words, if our approach was equivalent to bagging, it would only work for unstable vertex scoring functions. The fact that our procedure is effective even for PageRank, which is considered relatively stable \cite{ipsen2006mathematical,ng2001link}, corroborates the empirical findings that our method is capable of reducing bias in addition to variance. Rather than bagging, we think that our approach is more closely related to \textit{noise injection} techniques such as the \textit{noisy} and \textit{smoothed} bootstrap \cite{raviv1996bootstrapping,silverman1986density}, and to \textit{data augmentation} strategies. More generally, perturbing graphs could also be seen as a form of \textit{adversarial training} \cite{adversarial}.

\section{Discussion}

\begin{figure*}[ht!]
    \centering
        \setlength{\belowcaptionskip}{-4pt}
    \captionsetup{size=small}
	\includegraphics[width=0.4\textwidth]{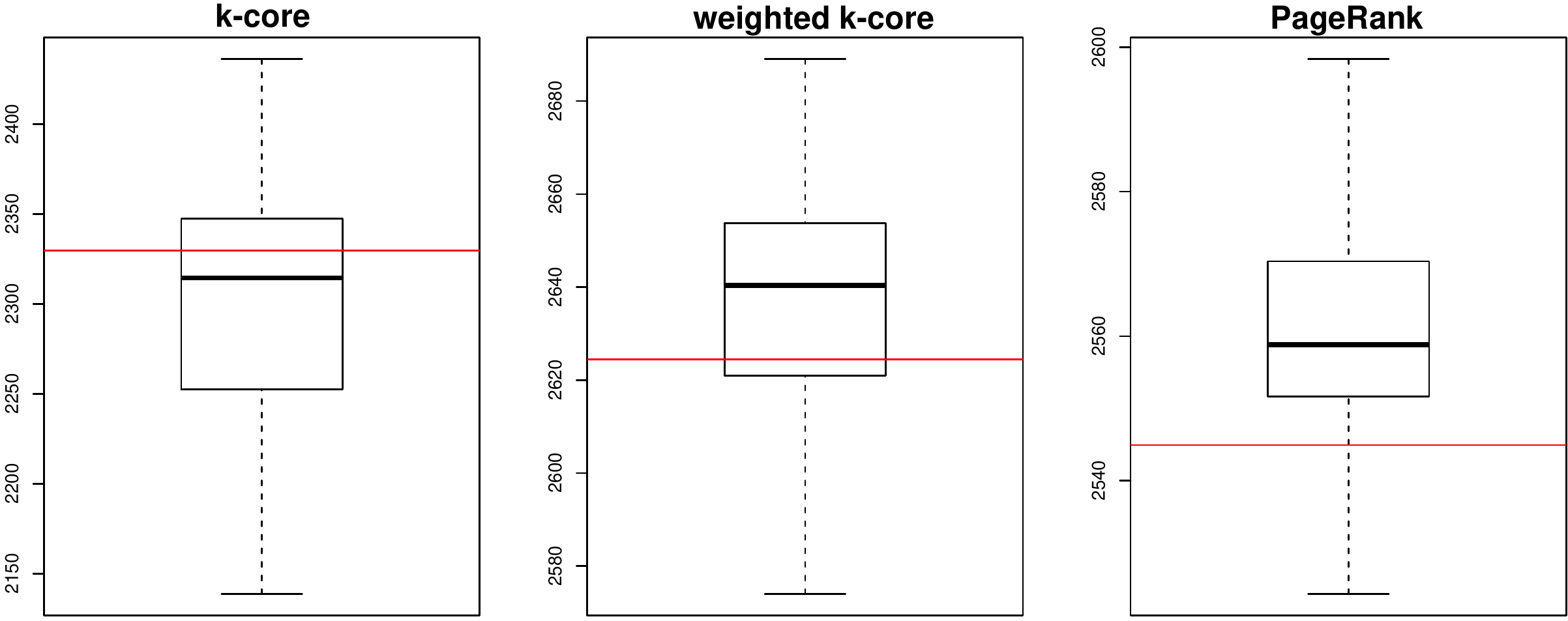}
	\hspace{0.5cm}
		\includegraphics[width=0.4\textwidth]{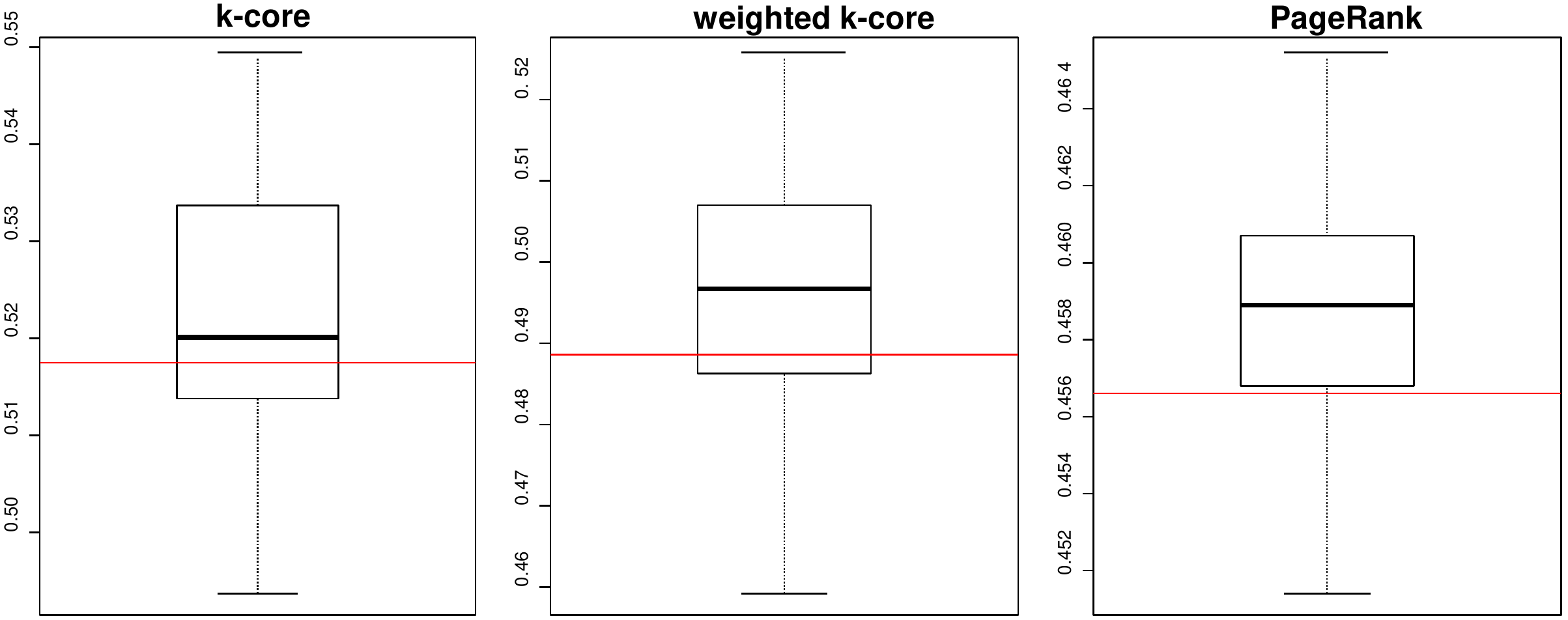}
	\caption{Left: distribution of epidemic severity for the 120 parameter combinations on EPINIONS. Right: distribution of macro-averaged F1-score for the 180 parameter combinations on Hulth2003. Horizontal lines indicate performance when using original scores. \label{fig:boxplots}}
\end{figure*}

\begin{table*}[ht!]
\small
\captionsetup{size=small}
\setlength{\abovecaptionskip}{-6pt}
\caption{Top 5 best P\&C parameters for each scoring function, in terms of number of nodes infected, on the social networks.}\label{table:best_social}
    \begin{center}                  
        \resizebox{0.9\textwidth}{!}{
\begin{tabular}{c|cccccccc||cccccccc||cccccccc}
  \hline
& $\delta_w$ & $\mathbb{G}$ & $M$ & $\varepsilon_d$ & $\varepsilon_a$ & \textbf{pr} & $cw$ & $cu$ & $\delta_w$ & $\mathbb{G}$ & $M$ & $\varepsilon_d$ & $\varepsilon_a$ & \textbf{cw} & $pr$ & $cu$ & $\delta_w$ & $\mathbb{G}$ & $M$ & $\varepsilon_d$ & $\varepsilon_a$ & \textbf{cu} & $pr$ & $cw$ \\ 
  \hline
\multirow{5}{*}{\rotatebox[origin=c]{90}{ENRON}} & 0 & $ER$ & 64 & 0.2 & 0.2 & \textbf{2454} & 2641 & 2443 & 0 & $CL$ & 64 & 0.2 & 0 & \textbf{2724} & 2225 & 2456 & 1 & $ER$ & 64 & 0.2 & 0.05 & \textbf{2538} & 2349 & 2666 \\ 
 & 1 & $ER$ & 16 & 0.2 & 0.2 & \textbf{2454} & 2613 & 2444 &   0 & $CL$ & 64 & 0.1 & 0.1 & \textbf{2718} & 2343 & 2415 &  1 & $ER$ & 64 & 0.2 & 0.1 & \textbf{2529} & 2419 & 2525 \\ 
 & 0 & $ER$ & 16 & 0.05 & 0.2 & \textbf{2447} & 2663 & 2434 &   1 & $CL$ & 64 & 0.2 & 0.2 & \textbf{2717} & 2360 & 2426 & 0 & $ER$ & 64 & 0.1 & 0.1 & \textbf{2521} & 2409 & 2675 \\ 
 & 1 & $ER$ & 64 & 0.2 & 0.2 & \textbf{2446} & 2684 & 2437 &  1 & $CL$ & 64 & 0.1 & 0.1 & \textbf{2710} & 2364 & 2445 & 0 & $ER$ & 64 & 0.05 & 0.05 & \textbf{2513} & 2422 & 2625 \\ 
 & 1 & $ER$ & 16 & 0.05 & 0.2 & \textbf{2445} & 2649 & 2474 &  1 & $ER$ & 64 & 0.05 & 0 & \textbf{2709} & 2333 & 2444 &   0 & $ER$ & 64 & 0.1 & 0.05 & \textbf{2508} & 2321 & 2551 \\ 
   \hline
\multirow{5}{*}{\rotatebox[origin=c]{90}{EPINIONS}} & 1 & $ER$ & 16 & 0.2 & 0.2 & \textbf{2598} & 2638 & 2391 & 0 & $ER$ & 64 & 0.1 & 0 & \textbf{2689} & 2553 & 2332 & 1 & $ER$ & 16 & 0.2 & 0.05 & \textbf{2436} & 2574 & 2653 \\
 & 1 & $CL$ & 64 & 0.2 & 0.05 & \textbf{2597} & 2654 & 2296 &  1 & $CL$ & 64 & 0.05 & 0.05 & \textbf{2673} & 2543 & 2246 & 1 & $ER$ & 16 & 0.2 & 0.1 & \textbf{2436} & 2581 & 2646 \\ 
 & 1 & $ER$ & 64 & 0.2 & 0.2 & \textbf{2596} & 2652 & 2388 &  0 & $ER$ & 64 & 0.2 & 0.05 & \textbf{2673} & 2552 & 2265 &  1 & $CL$ & 64 & 0.2 & 0 & \textbf{2423} & 2573 & 2667 \\ 
 & 1 & $ER$ & 16 & 0.1 & 0.2 & \textbf{2596} & 2631 & 2370 &  0 & $ER$ & 64 & 0.1 & 0.2 & \textbf{2672} & 2565 & 2324 &  1 & $ER$ & 16 & 0.2 & 0 & \textbf{2419} & 2557 & 2653 \\ 
 & 1 & $CL$ & 64 & 0.2 & 0.2 & \textbf{2593} & 2607 & 2310 &  1 & $CL$ & 64 & 0.1 & 0.1 & \textbf{2672} & 2573 & 2245 &  1 & $ER$ & 16 & 0.2 & 0.2 & \textbf{2391} & 2598 & 2638 \\ 
   \hline
\multirow{5}{*}{\rotatebox[origin=c]{90}{WIKIVOTE}} & 0 & $ER$ & 16 & 0.2 & 0.2 & \textbf{596} & 523 & 474 & 0 & $ER$ & 64 & 0.2 & 0 & \textbf{612} & 584 & 469 & 1 & $CL$ & 16 & 0.05 & 0.05 & \textbf{490} & 575 & 532 \\ 
 & 0 & $ER$ & 16 & 0 & 0.2 & \textbf{595} & 523 & 463 & 0 & $CL$ & 64 & 0.2 & 0.05 & \textbf{600} & 574 & 440 & 1 & $ER$ & 64 & 0.2 & 0 & \textbf{488} & 578 & 555 \\ 
  & 1 & $ER$ & 64 & 0.05 & 0.2 & \textbf{594} & 537 & 465 & 0 & $CL$ & 64 & 0.2 & 0.1 & \textbf{589} & 584 & 438 & 1 & $CL$ & 16 & 0.1 & 0.05 & \textbf{487} & 581 & 530 \\ 
  & 0 & $ER$ & 16 & 0.1 & 0.2 & \textbf{593} & 542 & 450 & 0 & $CL$ & 64 & 0.1 & 0.1 & \textbf{582} & 573 & 453 & 1 & $ER$ & 16 & 0.1 & 0 & \textbf{487} & 572 & 529 \\ 
  & 0 & $ER$ & 16 & 0.2 & 0.05 & \textbf{593} & 529 & 461 & 0 & $CL$ & 64 & 0.1 & 0.05 & \textbf{578} & 581 & 455 & 1 & $ER$ & 16 & 0.05 & 0 & \textbf{485} & 576 & 530 \\
  \hline
\end{tabular}
}

\end{center}
\end{table*}

\begin{table*}[ht!]
\captionsetup{size=small}
\setlength{\abovecaptionskip}{-6pt}
\setlength{\belowcaptionskip}{-6pt}
\caption{Top 5 best P\&C parameters for each scoring function, in terms of F1-score, on Hulth2003.}\label{table:best_nlp}
    \begin{center}                  
        \resizebox{0.9\textwidth}{!}{
\begin{tabular}{cccccccc||cccccccc||cccccccc}
  \hline
 $\delta_{w}$ & $\mathbb{G}$ & $M$ & $\varepsilon_d$ & $\varepsilon_a$ & \textbf{cu} & $cw$ & $pr$ &  $\delta_{w}$ & $\mathbb{G}$ & $M$ & $\varepsilon_d$ & $\varepsilon_a$ & $cu$ & \textbf{cw} & $pr$ &  $\delta_{w}$ & $\mathbb{G}$ & $M$ & $\varepsilon_d$ & $\varepsilon_a$ & $cu$ & $cw$ & \textbf{pr} \\ 
  \hline
1 & $ER$ &  8 & 0.1 & 0.3 & \textbf{54.88} & 50.67 & 45.99 & 0 & $ER$ & 96 & 0.0 & 0.3 & 54.85 & \textbf{52.50} & 45.99 & 1 & $CL$ & 32 & 0.0 & 0.3 & 50.94 & 45.33 & \textbf{46.75} \\ 
0 & $ER$ & 96 & 0.0 & 0.3 & \textbf{54.85} & 52.50 & 45.99 & 0 & $ER$ &  8 & 0.0 & 0.3 & 54.0.2 & \textbf{52.38} & 45.52 & 0 & $CL$ & 32 & 0.3 & 0.3 & 50.45 & 49.46 & \textbf{46.57} \\
1 & $ER$ & 96 & 0.3 & 0.3 & \textbf{54.78} & 51.76 & 45.65 & 0 & $ER$ &  8 & 0.1 & 0.3 & 54.69 & \textbf{52.38} & 46.06 & 1 & $CL$ &  8 & 0.0 & 0.3 & 51.86 & 46.99 & \textbf{46.53} \\ 
0 & $ER$ &  8 & 0.1 & 0.3 & \textbf{54.69} & 52.38 & 46.06 & 0 & $ER$ & 32 & 0.0 & 0.3 & 54.67 & \textbf{52.31} & 45.90 & 0 & $CL$ & 32 & 0.0 & 0.3 & 52.39 & 50.07 & \textbf{46.47} \\ 
0 & $ER$ & 32 & 0.0 & 0.3 & \textbf{54.67} & 52.31 & 45.90 & 0 & $ER$ & 96 & 0.1 & 0.3 & 54.62 & \textbf{51.78} & 46.20 & 0 & $CL$ & 32 & 0.2 & 0.3 & 52.22 & 50.49 & \textbf{46.45} \\
\hline
\end{tabular}
}
\end{center}
\end{table*}

\noindent \textbf{Importance of parameters}. As can be seen in Fig. \ref{fig:boxplots}, most -but not all- parameter combinations return scores that allow the identification of better spreaders than the original scores. This suggests that while P\&C is relatively robust to the choice of parameter values, some optimization is necessary to get the most out of the procedure. Tables \ref{table:best_social} and \ref{table:best_nlp}) support this claim, by clearly showing that there is no single best combination of parameters across networks and vertex scoring functions.
Nonetheless, depending on the graph and/or scoring function, some parameters seem more important than others. For instance, for $cu$, selecting edges uniformly at random (ER model) during perturbation tends to work better than selecting edges in a biased way (CL model), which is consistent with the fact that $cu$ ignores edge weights, unlike $cw$ and $pr$. For $cw$, generating many perturbed versions of the original network (large values of $M$) also seems advantageous. Further research and analysis should help us understand what are the crucial parameters for different settings (graph type, size, density, diameter, scoring function...) and what are good initial values for them, reducing the need for parameter tuning.

\noindent \textbf{P\&C improves even the performance of PageRank}. This was unexpected, as PageRank is believed to be stable to edge-based perturbations \cite{ipsen2006mathematical,ng2001link}. The implication could be that our P\&C procedure is beneficial to any node scoring function, not only unstable ones, or that PageRank features some level of instability. While our theoretical analysis supports the former implication (see section \ref{sec:theory}), the latter is a legitimate possibility as well. It has indeed been suggested that the stability of PageRank depends on the network topology. E.g., PageRank is more stable for scale-free graphs like the Web than for random networks \cite{ghoshal2011ranking}.

\noindent \textbf{For word networks, adding edges is beneficial}.
Interestingly, for word co-occurrence networks, adding edges or incrementing the weights of already existing edges seems much more important than deleting edges ($\varepsilon_a \geq \varepsilon_d$), regardless of the scoring function.
This is equivalent to copying and pasting words from/to the input text, which can be seen as a form of \textit{data augmentation}. It could also be interpreted as having a sliding window of stochastic size featuring an additional \textit{masking} mechanism such that edges are drawn between a subset only of the words in each instantiation of the window. Data augmentation and stochastic windows were proven very beneficial in Computer Vision and NLP \cite{krizhevsky2012imagenet,mikolov2013efficient}, so this could explain why $\varepsilon_a \geq \varepsilon_d$ works well with word networks.

\vspace{-0.3cm}

\section{Conclusion}\label{sec:conc}
We proposed what is, to the best of our knowledge, the first application of the Perturb and Combine (P\&C) strategy to graphs. Experiments on real-world networks reveal that P\&C improves influential spreader detection performance for 3 vertex scoring functions. The P\&C scores can be obtained at little extra cost via parallelization. Finally, a theoretical analysis sheds light on why our P\&C strategy is effective, and suggests that it works mainly by reducing bias. This implies that P\&C could be effective even for stable algorithms.

\vspace{-0.15cm}

\section{Acknowledgements}
We thank the four anonymous reviewers for their feedback.

\bibliographystyle{IEEEtran}
\bibliography{paper}

\section*{Appendix A: proof of Equation \ref{eq:squared_error}}\label{app:proof}

\noindent We start by summing the squared bias and the variance (bias and variance have been defined in Equations \ref{eq:bias} and \ref{eq:var}):

\vspace{-0.35cm}

\begin{equation}
\mathrm{bias}^2\big[\Rest\big] + \mathrm{var}\big[\Rest\big] = \big( 1- \mathbb{E}\big[\mathrm{MET}\big] \big)^2 +  \mathbb{E}\big[\big(\mathrm{MET} - \mathbb{E}\big[\mathrm{MET}\big]\big)^2\big]
\end{equation}

\noindent Developing the left term of the right-hand side gives:
\begin{equation}
\mathrm{LRHS} = 1 - 2\mathbb{E}\big[\mathrm{MET}\big] + \mathbb{E}^2\big[\mathrm{MET}\big]
\end{equation}

\noindent Since the expected value of a sum of random variables is equal to the sum of their expected values, the expected value of a constant is equal to that constant, and for all random variable $X$ and $\lambda \in \mathbb{R}$, $\mathbb{E}\big[\lambda X \big]=\lambda\mathbb{E}\big[X\big]$, developing the right term of the right-hand side gives: \\
\begin{align}
\mathrm{RRHS} & = \mathbb{E}\big[\mathrm{MET}^2 - 2\mathbb{E}\big[\mathrm{MET}\big]\mathrm{MET} + \mathbb{E}^2\big[\mathrm{MET}\big] \big] \\ & =
\mathbb{E}\big[\mathrm{MET}^2\big] - 2 \mathbb{E}^2\big[\mathrm{MET}\big] + \mathbb{E}^2\big[\mathrm{MET}\big]
\end{align}

\noindent We thus get:

\begin{align}
\mathrm{bias}^2\big[\Rest\big] + \mathrm{var}\big[\Rest\big] & = \mathrm{LRHS} + \mathrm{RRHS} \\
& = 1 -2\mathbb{E}\big[\mathrm{MET}\big] + \mathbb{E}\big[\mathrm{MET}^2\big] \\
& = \mathbb{E}\big[\big(\mathrm{MET}-1\big)^2\big]
\end{align}

\noindent Which gives us Eq. \ref{eq:squared_error}. \QEDB

\newpage

\onecolumn

\section*{Appendix B: Example}\label{app:motiv_ex}

Note that we adopt here the theoretical framework introduced in Section \ref{sec:theory}. In this framework, we consider \textit{node scoring functions}. Such functions map each vertex to a score and as such provide a ranking of all the nodes in the network. This ranking is compared to the true ranking provided by the SIR model. Regardless of the node scoring function used, in this framework, a node with a score of 2 is considered twice as influential than a node with a score of 1.

\begin{figure}[h]
\setlength\tabcolsep{3pt}
\renewcommand{\arraystretch}{0.42}
    \centering
\captionsetup{size=small}
	\includegraphics[width=0.48\textwidth]{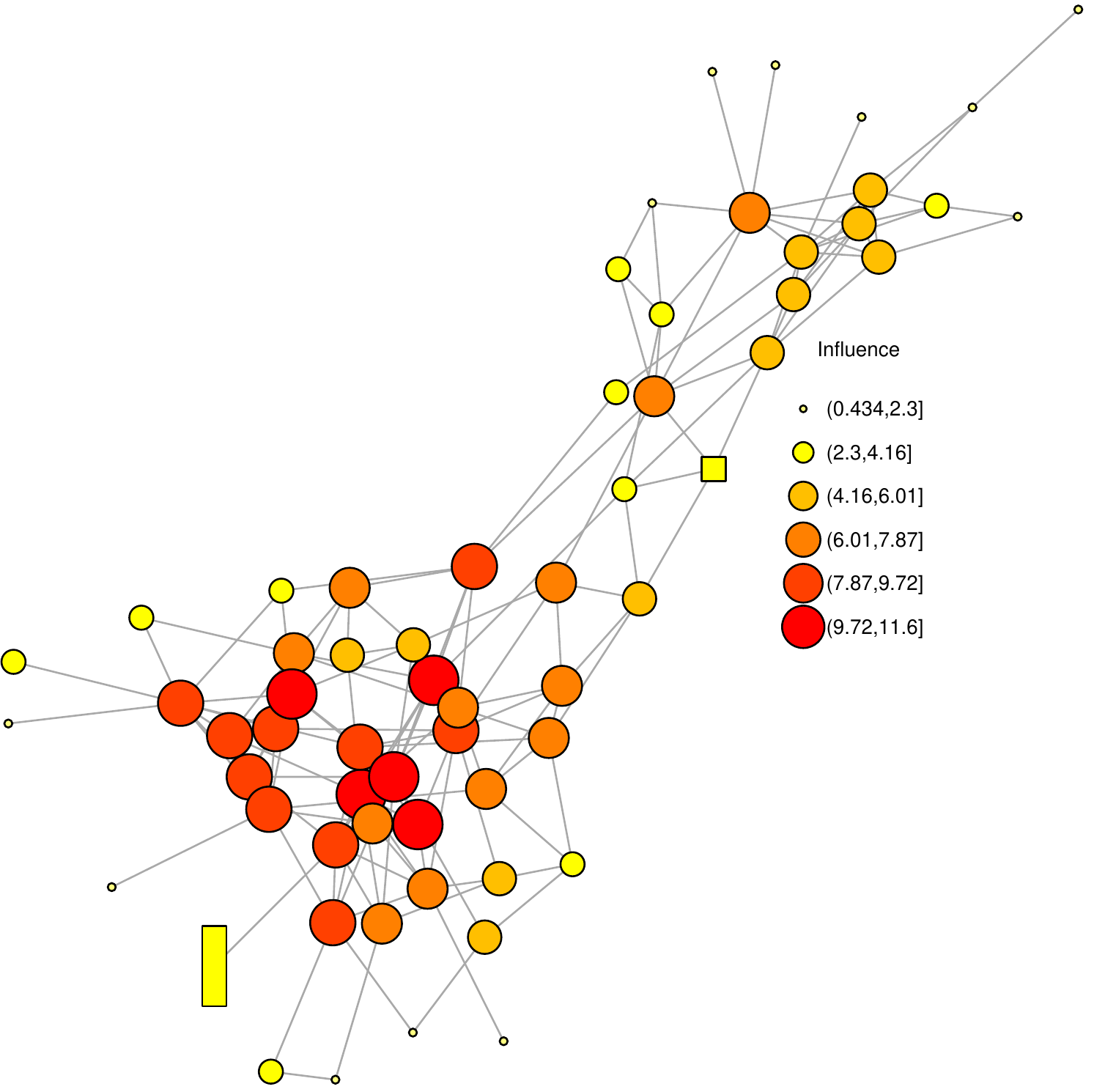}
		\caption{SIR-based spreading influence of the nodes in a toy network. The table below shows the original (`orig.') and P\&C scores of the square and rectangle nodes, for $k$-core. The P\&C score of a node is the mean of its scores in each of the 8 perturbed graphs. Truth denotes the SIR influence. Note that scores and influence are in different units, only ratios (i.e., rankings) matter. \label{fig:truth_dolphins}}
	\scalebox{0.99}{
\begin{tabular}{rccccccccccc}
 & orig. & \multicolumn{8}{c}{perturbed} & P\&C & truth\\ 
  \hline
square & \textbf{4} & 3 & 3 & 3 & 2 & 1 & 2 & 3 & 2 & \textbf{2.38} & 4.09 \\ 
  rectangle & \textbf{1} & 1 & 2 & 3 & 1 & 2 & 1 & 1 & 2 & \textbf{1.62} & 2.76 \\ 
   \hline
\end{tabular}

}
	
\end{figure}

Look at the square and rectangle nodes in Fig. \ref{fig:truth_dolphins}. In the original graph, the square node is a member of the main core ($k=4$), but a quick visual inspection reveals that this node does not lie in the most central part of the network and is not strongly attached to the main core. With a degree of only 4, the square node is actually one of the weakest members of the main core, i.e., removing only one of its connections would suffice in decreasing its score.

\vspace{-1cm}

\noindent On the other hand, the rectangle node is part of the $1$-core in the original network. This low score does not reflect the fact that this node has direct access to the most central part of the network through its single connection. Should an epidemic be triggered from that node, it would probably be more severe than its low score suggests. To sum up, based on the original scores, the square node is 4 times more influential than the rectangle node. This is far from reality, as the ratio of the true influence scores is only $4.09/2.76=1.48$.

Looking at the scores obtained by the square and rectangle nodes in 8 slightly perturbed versions of the original network (a few edges added/deleted at random), we can observe that the rectangle node gets higher scores in most perturbed graphs, whereas the square node gets lower scores most of the time. Using the average of these 8 scores instead of the original scores is much closer to the true ratio: $2.38/1.62=1.47$.

Note that since we both add and remove edges in our perturbation strategy, the square node could get higher scores, and the rectangle node could get lower scores (e.g., it could get disconnected, and get a score of 0). There are no constraints on the values node scores can take on. The fact that the rectangle node gets higher scores in most perturbed graphs, whereas the square node gets lower scores in most of them, is truly reflective of the positions of the nodes in the graph.

\section*{Appendix C: illustration of the perturbation strategy}\label{app:perturbation_illustration}

\section*{Appendix D: word co-occurrence network example}\label{app:gow}

\end{document}